# LOCALIZATION BASED SEQUENTIAL GROUPING FOR CONTINUOUS SPEECH SEPARATION

*Zhong-Qiu Wang[1] and DeLiang Wang[2]*

[1]Mitsubishi Electric Research Laboratories (MERL), USA

[2]Department of Computer Science and Engineering, The Ohio State University, USA

wang.zhongqiu41@gmail.com, dwang@cse.ohio-state.edu

## ABSTRACT

This study investigates robust speaker localization for continuous speech separation and speaker diarization, where we use speaker directions to group non-contiguous segments of the same speaker. Assuming that speakers do not move and are located in different directions, the direction of arrival (DOA) information provides an informative cue for accurate sequential grouping and speaker diarization. Our system is block-online in the following sense. Given a block of frames with at most two speakers, we apply a two-speaker separation model to separate (and enhance) the speakers, estimate the DOA of each separated speaker, and group the separation results across blocks based on the DOA estimates. Speaker diarization and speaker-attributed speech recognition results on the LibriCSS corpus demonstrate the effectiveness of the proposed algorithm.

## 1. INTRODUCTION

Riding on deep learning, dramatic progress has been made in single- and multi-microphone speaker diarization in noisy-reverberant conditions with overlapped speech [1], [2]. Many diarization studies address the vanilla problem of *who spoke when*, and consider speech separation and automatic speech recognition (ASR) as downstream tasks [3]–[6]. In the CHiME-5 and 6 challenges [7], [8], two popular approaches are the guided source separation technique [9], which performs spatial clustering and beamforming based separation using oracle diarization annotations, and target-speaker voice activity detection [10], which trains a DNN on a combination of spectral features and the i-vector of each speaker to directly predict the activity of each speaker at each frame. Another popular approach first performs continuous speech separation [11]–[16], which produces multiple monaural output streams, each with no speaker overlap. Then spectral features are extracted from the monaural, separated signals for diarization. Intuitively, if the separation result is sufficiently accurate, later diarization and recognition would be improved [17]. In this study, instead of using spectral information, we investigate the use of spatial information for speaker diarization. The motivation is that, if speakers are not spatially overlapped and do not move, the DOA information of each speaker provides an informative cue for diarization.

This research was supported in part by an NSF grant (ECCS-1808932) and the Ohio Supercomputer Center.

In multi-microphone speaker diarization, using DOA information for diarization has been explored in early diarization studies [1], but seldomly in recent deep learning based studies. In [18], a GMM-HMM system is built on GCC-PHAT features for location based speaker diarization. In [19], a time difference classsification system is built for DOA based diarization. In [20], GMM-HMM based joint spectral and spatial modeling is conducted for diarization. In [21], [1], Anguera *et al.* investiagte GCC-PHAT based time difference of arrival (TDOA) for delay-and-sum beamforming, and use the TDOA estimates together with the spectral features extracted from the beamforming results for diarization. In [22], GCC-PHAT based DOA estimation is performed after dereverberation and the DOA results across frames are clustered for diarization. Although a lot of research has been conducted in using DOA information to improve diarization, we point out that many studies extract DOA information directly from noisy-reverberant multi-speaker mixtures (or from enhanced or separated mixtures that are not sufficiently accurate), which usually lead to inaccurate time-delay features and DOA estimation. For example, the summated GCC-PHAT coefficients would exhibit spurious and broad peaks in noisy-reverberant conditions, and multiple peaks if there are competing speakers and directional noises [23].

Recently, the performance of single- and multi-microphone speech enhancement, dereverberation and speaker separation [15], [24]–[31] has been dramatically improved using deep learning. Such improved enhancement or separation results can benefit DOA estimation. In [32], [23], estimated time-frequency (T-F) masks of target speakers are utilized to identify T-F units with cleaner phase for DOA estimation, producing dramatic improvement over the vanilla GCC-PHAT algorithm. In this context, this study addresses the problem of *who spoke what at when and where*, where we first use a state-of-the-art DNN model for block-online separation, and then uses the separation results to compute the DOA of each speaker at each block. The DOA estimates are then clustered across blocks to get the diarization and separation results of each speaker for ASR.

The rest of this paper is organized as follows. We present the proposed algorithm Section 2, and experimental setup and evaluation results in Section 3 and 4. Section 5 concludes this paper.

## 2. PROPOSED ALGORITHMS

Our system is block-online. In each short processing block, we assume that there are at most $C$ (=2 in this study) speakers. Within each block, the physical model of a $P$-microphone mixture can be formulated in the short-time Fourier transform (STFT) domain as

$$Y(t,f) = \sum_{c=1}^{C} X(c,t,f) + N(t,f) \\ = \sum_{c=1}^{C} \big(S(c,t,f) + H(c,t,f)\big) + N(t,f), \quad (1)$$

where $Y(t,f)$, $N(t,f)$, $S(c,t,f)$, $H(c,t,f)$ and $X(c,t,f) \in \mathbb{C}^P$ respectively denote the complex STFT vectors of the received mixture, reverberant noise, and direct-path signal, early reflections plus late reverberation, and reverberant image of speaker $c$ at time $t$ and frequency $f$. In the following sections, when dropping $t$ and $f$ from the notation, we refer to the corresponding spectrogram. For example, $Y_q$ denotes the spectrogram of the mixture at microphone $q$, and $S_q(c)$ denotes that of speaker $c$. We aim at estimating $S_q(c)$ for each source at the reference microphone based on the multi-channel input $Y$. Our study assumes a common uniform circular array geometry and that the same array is used for training and testing. The first microphone on the circle is always considered as the reference microphone, i.e. $q = 1$.

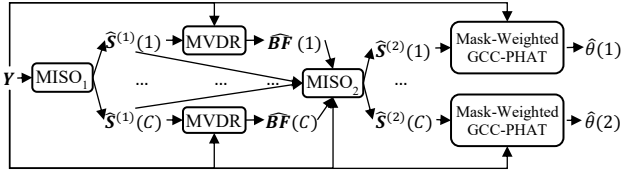

Figure 1. MISO-BF-MISO for block-online separation and localization.

Table 1. Conversion table from speaker counting results to VAD results.

| Speaker counting results | Stream 1 active? | Stream 2 active? |
|---|---|---|
| 0 | No | No |
| 1 | Yes if $\sum_f \left|\hat{S}_q^{(2)}(1,t,f)\right|^2 \geq \sum_f \left|\hat{S}_q^{(2)}(2,t,f)\right|^2$ else No | Yes if $\sum_f \left|\hat{S}_q^{(2)}(1,t,f)\right|^2 < \sum_f \left|\hat{S}_q^{(2)}(2,t,f)\right|^2$ else No |
| 2 | Yes | Yes |

At each block, we perform separation (and enhancement) and count the number of speakers at each frame. Based on the separation and counting results, we localize each speaker. Next, we group the separation results across blocks based on the localization results, and feed the grouped separation results into an ASR backend for recognition. The rest of this section details each step.

### 2.1. MISO-BF-MISO for Block-Online Separation

We employ a state-of-art speaker separation model, MISO-BF-MISO [33], for separation at each block. See Figure 1 for an illustration. It contains two multi-microphone input and single-microphone output (MISO) networks, with a time-invariant minimum variance distortionless response (MVDR) module in between. Both networks are trained using multi-microphone complex spectral mapping [34], [33], where we predict the real and imaginary (RI) components of target speech at a reference microphone from the RI components of the stacked multi-channel input signals. The first network is trained using utterance-wise permutation invariant training [26] to estimate the direct-path signal of each speaker at each microphone, denoted as $\hat{S}_q^{(1)}(c)$, where the superscript indicates that it is produced by the first DNN. The target estimates are then utilized to compute spatial covariance matrices for MVDR beamforming. The second MISO network takes in the outputs of the first network and the beamforming results, denoted as $\widehat{BF}(c)$ in Figure 1, to enhance each target speaker. The output is denoted as $\hat{S}_q^{(2)}(c)$. More details can be found in [33].

### 2.2. Frame-Wise Speaker Counting

At each block, the separation module produces one stream of outputs for each speaker. For each speaker stream, we need to identify frames with active speech for accurate diarization. Based on the separation results, one simple but error-prone way is to set a frame-level energy threshold for voice activity detection (VAD). Rather than doing this, we train a MISO based speaker counting network to predict the number of speakers at each frame. The input features are the stacked RI components of the mixture, plus the magnitude features at the reference microphone. As each block is assumed to have at most two speakers, we perform three-class classification (i.e. 0, 1 and 2 speakers) for frame-wise speaker counting. The model is trained using cross-entropy. On our simulated reverberant two-speaker mixtures, introduced later in Section 3, the accuracy of speaker counting is around 97%, which is reasonably accurate.

Note that the speaker counting result at each frame is either 0, 1 or 2 speakers, while for each frame in each of the two separated streams, we need to determine whether or not there is active speech (i.e. 0 or 1 speaker). We do the conversion following the rules in Table 1. We denote the binary VAD results as $\hat{v}(c,t)$. It is used to compute the boundary of each speech segment for source $c$. It will also be used in later DOA estimation.

### 2.3. Mask-Weighted GCC-PHAT for DOA Estimation

In meeting scenarios, speakers are assumed sitting in their chairs and do not move too much. In addition, the speakers are assumed not spatially overlapped with respect to the microphone array. In such a case, the DOA information of the speakers provides an informative cue for accurate diarization. The challenge here is how to perform accurate DOA estimation for each speaker given a noisy-reverberant recording with overlapped speech.

In our recent work on localizing a single target speaker in noisy-reverberant environments [23], we found that the target speaker can be localized very accurately using deep learning based T-F masking. The key idea is to use a DNN to identify T-F units dominated by the target speaker and only use these T-F units for localization, as such T-F units contain cleaner phase informative for accurate localization. The T-F masks can be accurately estimated by using a DNN trained to estimate, say, the ideal ratio mask [27]. In other words, as long as we have a separation result of a target speaker, we can leverage it to localize that speaker. In this study, we extend this localization technique to multi-speaker localization in noisy-reverberant conditions.

Assuming a known array geometry, we first compute the vanilla GCC-PHAT coefficients [35]

$$G_{p,p'}(t,f,k) \\ = \text{Real}\left\{\frac{Y_p(t,f)Y_{p'}(t,f)^\text{H}}{|Y_p(t,f)||Y_{p'}(t,f)^\text{H}|}e^{-j2\pi\frac{f}{N}f_s\tau_{p,p'}(k)}\right\} \\ = \cos\left(\angle Y_p(t,f) - \angle Y_{p'}(t,f) - 2\pi\frac{f}{N}f_s\tau_{p,p'}(k)\right), \quad (2)$$

where $p$ and $p' \in \{1,\dots,P\}$ are the indices of a pair of microphones, Real($\cdot$) extracts the real component, $(\cdot)^\text{H}$ computes complex conjugate, $N$ is the number of discrete-time Fourier transform

(DFT) frequencies, $f_s$ the sampling rate, $k$ a hypothesized direction, and $\tau_{p,p'}(k)$ the hypothesized time delay between microphone $p$ and $p'$ if the speaker is in the hypothesized direction $k$. $f$ indexes from 0 to $N/2$. GCC-PHAT coefficients essentially measure the cosine distance between the observed IPD, $\angle Y_p(t,f) - \angle Y_{p'}(t,f)$, and hypothesized IPD, $2\pi \frac{f}{N} f_s \tau_{p,p'}(k)$, at each T-F unit. If the distance is small, it means that the dominant source at the T-F unit is from the hypothesized direction and from other directions otherwise. The GCC-PHAT coefficients are then summated over all the microphone pairs and over all the T-F units within the block. The direction $k$ producing the highest summation is considered as the estimated direction.

Although, in conditions with mild reverberation, GCC-PHAT shows reasonable performance, strong directional noises or interference speakers, and diffuse noises or room reverberation would broaden the peaks and create spurious peaks in the summated GCC-PHAT coefficients, as interferences are typically in the other directions. To sharpen the target peak and suppress other spurious peaks, based on the separation result $\hat{S}_q^{(2)}(c)$ we identify T-F units dominated by source $c$ and only use them for localization, by using a weighting mechanism [23]

$$\widehat{MG}_{p,p'}(c,t,f,k) = \widehat{M}_{p,p'}(c,t,f) \frac{G_{p,p'}(t,f,\tau_{p,p'}(k)) + 1}{2}, \quad (3)$$

where we add the GCC-PHAT coefficient by one and divide it by two to normalize it into the range [0,1], and $\widehat{M}_{p,p'}(c,t,f)$ is a weight denoting the importance of the T-F for localizing speaker $c$. The weight is computed as a product of the estimated ratio mask at each of the two microphones so that higher weight is given to T-F units where speaker $c$ is dominant at both microphones

$$\widehat{M}_{p,p'}(c,t,f) = \widehat{M}_p(c,t,f) \widehat{M}_{p'}(c,t,f) \quad (4)$$
$$\widehat{M}_p(c) = |\hat{S}_p(c)|/(|\hat{S}_p(c)| + |Y_p - \hat{S}_p(c)|) \quad (5)$$
$$\widehat{M}_{p'}(c) = |\hat{S}_{p'}(c)|/(|\hat{S}_{p'}(c)| + |Y_{p'} - \hat{S}_{p'}(c)|). \quad (6)$$

Similarly to the GCC-PHAT coefficients, the mask-weighted GCC-PHAT coefficients are then summated together over all the microphone pairs and over all the T-F units within each block. The hypothesized direction producing the largest summation is considered as the estimated direction. Mathematically,

$$\widehat{SMG}(c,k) = \sum_{(p,p') \in \Omega} \sum_t \hat{v}(c,t) \sum_{f=1}^{N/2} \widehat{MG}_{p,p'}(c,t,f,k), \quad (7)$$

where $\Omega$ contains all the microphone pairs, and

$$\hat{\theta}(c) = \mathrm{argmax}_k \widehat{SMG}(c,k). \quad (8)$$

We only perform localization for non-silent streams at each block. This can be determined based on the VAD result $\hat{v}(c,t)$.

### 2.4. DOA Based Sequential Grouping

Given the DOA estimate of each speaker at each block, we group the separation results across all the blocks according to their DOA estimates. As an initial step towards location based diarization, this paper uses a simple block-online algorithm to create and update clusters (see Algorithm 1). Basically, we maintain a list for existing clusters, each including a DOA estimate. For each stream in a new block, we merge it with an existing cluster if their DOA

```
𝒳 = [];
For each block b do
    For c = 1: C do
        If 𝒳 is not empty do
            - Find from 𝒳 the element with its angle closest to θ̂(c) and
              denote its index as i;
            - If θ̂(c) is within 5° from 𝒳[i][0] do
                𝒳[i][1] += SMG(c, k);
                𝒳[i][0] = argmax 𝒳[i][1];
                Assign stream c of block b to cluster #i;
            - Else
                Assign stream c of block b to cluster #length(𝒳);
                𝒳. append([θ̂(c), SMG(c, k)]);
            - End
        Else
            - Assign stream c of block b to cluster #length(𝒳);
            - 𝒳. append([θ̂(c), SMG(c, k)]);
        End
    End
End
```
Algorithm 1. Pseudocode for block-online DOA-based sequential grouping.

estimates are within 5° to each other; and if it is 5° away from all the existing clusters, we create a new cluster.

We emphasize that our diarization system is block-online and we do not need to train a DNN model specifically for diarization.

## 3. EXPERIMENTAL SETUP

We validate the proposed algorithms on the LibriCSS dataset [14], which contains ten hours of conversational speech data recorded by playing LibriSpeech signals through loud speakers in reverberant rooms. The task is to perform conversational speech recognition in reverberant conditions with a wide range of speaker overlaps. There are ten one-hour sessions, each consisting of six ten-minute mini-sessions with speaker overlap ratios spanning from 0% to 40%, including 0S (no overlap with short inter-utterance silence between 0.1 and 0.5 s), 0L (no overlap with long inter-utterance silence between 2.9 and 3.0 s), and 10%, 20%, 30% and 40% overlaps. The recording device has seven microphones, with six of them uniformly spaced on a circle with a 4.25 cm radius, and one at the circle center. The speaker-to-array distance ranges from 33 to 409 cm. For speaker diarization and recognition, we consider **session-wise evaluation**, where each ten-minute mini-session signal is used for evaluation, and **segment-wise evaluation**, where each mini-session signal is pre-segmented by the authors of LibriCSS [14] into 5- to 120-second long segments, each with 2 to 10 utterances from up to 8 speakers.

Diarization performance is measured using diarization error rates (DER). We report ASR performance using concatenated minimum-permutation word error rates (cpWER) [8]. It is computed by concatenating all the utterances of each speaker in the hypothesis and reference, scoring all speaker pairs, and finding the permutation that produces the best WER. Note that the estimated number of speakers, $A$, could be different from the actual number of speakers, $B$. When $A < B$, we align the $A$ hypotheses to the references and consider all the rest $B - A$ references to produce deletion errors. When $A > B$, we can only align $B$ hypotheses with the references and compute the WER. We do not score the rest $A - B$ hypotheses. Note that when $A > B$, it is likely that some speakers are splitted into multiple output streams. It is fine if we do not score the rest $A - B$ hypotheses, as the hypothesis aligned to a splitted speaker is not produced by using all the speech of that speaker for recognition. As a result, the error of splitting a speaker into multiple streams would still have an influence on the final cpWER, even if the rest $A - B$ hypotheses are not scored.

Table 2. DER (%) and cpWER (%) on LibriCSS (Segment-Wise Evaluation).

| Separation Approaches | Diarization Approaches | DER | | | | | | | cpWER | | | | | | |
|---|---|---|---|---|---|---|---|---|---|---|---|---|---|---|---|
| | | Overlap Ratio (%) | | | | | | Avg. | Overlap Ratio (%) | | | | | | Avg. |
| | | 0S | 0L | 10 | 20 | 30 | 40 | | 0S | 0L | 10 | 20 | 30 | 40 | |
| Unprocessed | Offline x-vector+SC [15] | 22.38 | 15.91 | 22.06 | 27.45 | 32.97 | 35.30 | 26.76 | 35.21 | 32.07 | 42.27 | 51.36 | 59.98 | 61.66 | 47.09 |
| Mask-Based MVDR [14] | | 24.58 | 21.17 | 22.12 | 26.31 | 27.16 | 24.91 | 24.49 | 36.13 | 35.43 | 38.35 | 44.07 | 45.60 | 42.55 | 40.36 |
| Multi-Frame MCWF [16] | | 22.21 | 19.80 | 22.00 | 24.98 | 27.19 | 23.27 | 23.42 | 34.14 | 30.93 | 33.50 | 37.52 | 41.30 | 32.78 | 35.03 |
| MISO-BF-MISO [33] | | 19.89 | 16.56 | 17.52 | 18.34 | 21.95 | 18.07 | 18.72 | 28.84 | 26.44 | 29.33 | 32.69 | 38.72 | 33.10 | 31.52 |
| MISO-BF-MISO [33] | Block-Online DOA Based | 11.08 | 10.48 | 10.27 | 11.07 | 11.72 | 13.63 | 11.48 | 9.84 | 9.24 | 10.08 | 13.04 | 14.62 | 17.58 | 12.4 |

Table 3. DER (%) and cpWER (%) on LibriCSS (Session-Wise Evaluation).

| Separation Approaches | Diarization Approaches | DER | | | | | | | cpWER | | | | | | |
|---|---|---|---|---|---|---|---|---|---|---|---|---|---|---|---|
| | | Overlap Ratio (%) | | | | | | Avg. | Overlap Ratio (%) | | | | | | Avg. |
| | | 0S | 0L | 10 | 20 | 30 | 40 | | 0S | 0L | 10 | 20 | 30 | 40 | |
| Unprocessed | Offline x-vector+SC [15] | 9.27 | **4.98** | 11.18 | 16.78 | 22.00 | 25.70 | 15.89 | 12.72 | 12.74 | 21.25 | 30.73 | 37.29 | 44.91 | 26.61 |
| Mask-Based MVDR [14] | | 12.62 | 11.31 | 12.04 | 14.31 | 16.38 | 15.92 | 13.97 | 18.07 | 16.14 | 18.47 | 21.41 | 25.34 | 26.85 | 21.05 |
| Multi-Frame MCWF [16] | | 10.20 | 11.80 | 13.84 | 15.76 | 18.69 | 17.33 | 15.04 | 14.43 | 12.18 | 14.76 | 15.86 | 18.06 | 18.28 | 15.6 |
| MISO-BF-MISO [33] | | **9.20** | 7.09 | **7.53** | **7.19** | **10.11** | **8.83** | **8.33** | 10.84 | 9.24 | 11.05 | **12.27** | 16.39 | **17.42** | **12.87** |
| MISO-BF-MISO [33] | Block-Online DOA Based | 11.95 | 10.69 | 11.25 | 12.22 | 13.04 | 14.31 | 12.36 | **10.18** | **9.03** | **10.89** | 13.60 | **16.14** | 18.05 | 12.98 |

Since LibriCSS only has real-recorded testing data, we simulate our training and validation data to train the separation models. Our training data includes 76,750 (~129 hours) seven-channel two-speaker mixtures with mild room reverberation and weak stationary noises. The clean speech signals are sampled from the *train-clean-{100,360}* set of LibriSpeech. Assuming the array geometry of the LibriCSS recording device, we simulate seven-channel RIRs based on an RIR generator [36]. The reverberation time is sampled from the range [0.2,0.6] s. The average speaker-to-array distance is sampled from the range [0.75,2.5] m. The average direct-to-reverberation energy ratio of the RIRs is −0.3 dB with 3.9 dB standard deviation. The two speaker angles are sampled from $[-\pi, +\pi]$ and ensured to be at least $10°$ apart. The energy level between the two speakers is drawn from the range $[-7,7]$ dB. For each reverberant two-talker mixture, we sample an air conditioning noise from the REVERB corpus. The SNR between the anechoic two-speaker mixture and the noise is sampled from the range [10,30] dB. The labels for training the frame-wise speaker counting module are obtained by first applying a pre-trained voice activity detector [37] to the spatialized anechoic signal of each speaker at the reference microphone, and then combining the VAD results to get the number of speakers at each frame.

Following [14], [38], the run-time block size is set to 2.424 s, and block shift is set to 1.2 s. The hypothesized direction $k$ in Eq. (2) is enumerated from $-180°$ to $180°$, at a step of $1°$.

We use two publicly-available multi-channel separation results[1] of LibriCSS for comparison. The first one [39], [14], provided by the authors of LibriCSS, is based on T-F masking based MVDR with additional gain adjustment. It uses 2.4 s block size and 1.6 s block overlap, and assumes that there are at most two speakers at each block. The other one [16] is obtained by iteratively performing sequential multi-frame multi-channel Wiener filter (MCWF) and post-filtering. It uses 8 s block size and 4 s block overlap, and assumes that there are at most three speakers in each block.

We use the Kaldi recipe[2] [15] to build the diarization and recognition backends for LibriCSS. We use the TDNN-F model used in [15] for recognition. It is trained on Librispeech and fine-tuned on a reverberated version of Librispeech. An x-vector extractor is built using the VoxCeleb data with simulated RIRs [40], [15]. For each segment in each separated speaker stream, an x-vector is extracted and used for offline spectral clustering based diarization [15]. We denote this diarization approach as **x-vector+SC**.

## 4. EVALUATION RESULTS

Table 2 and Table 3 respectively report the DER and cpWER results on the segment- and session-wise evaluations. When using x-vector+SC for diarization, MISO-BF-MISO shows better DER and cpWER than the other two separation systems (for example in the segment-wise case, 18.72% vs. 24.49% and 23.42% in DER, and 31.52% vs. 40.36% and 35.03% in cpWER), indicating its strong separation performance. When using the proposed block-online DOA based system to perform diarization on the MISO-BF-MISO separation results, we observe clearly better DER and cpWER results than the offline x-vector+SC system in the segment-wise case (11.48% vs. 18.72% in DER and 12.4% vs. 31.52% in cpWER), while in the session-wise case worse DER is observed (12.36% vs. 8.33%) and comparable cpWER is obtained (12.98% and 12.87%). This is likely because longer recordings in the session-wise case can have more single-speaker segments than in the segment-wise case. Such segments can lead to more reliable speaker embedding centroids and a more accurate estimated number of clusters when modelled through offline clustering. These results demonstrate the effectiveness of using DNN-estimated DOA information for diarization, and indicate the strong potential of combining spectral and DOA information for diarization.

## 5. CONCLUSION

We have proposed for continuous speech separation a sequential grouping technique using deep learning based speaker separation and localization. Evaluation results on LibriCSS suggest that DOA information produced by using DNN-estimated target speech provides an informative cue for speaker diarization. Our future research will improve the block-online clustering module, and integrate spatial and spectral cues for diarization.

---
[1]Available online at https://zenodo.org/record/4415163/#.YIX-sRP0nNA
[2]Available online at https://github.com/kaldi-asr/kaldi/tree/master/egs/libri_css